\documentclass[conference]{IEEEtran}
\IEEEoverridecommandlockouts
\usepackage{cite}
\usepackage{amssymb,amsfonts}
\usepackage[fleqn]{amsmath}
\usepackage{array}
\usepackage{algorithmic}
\usepackage[dvipdfmx]{graphicx}
\usepackage{textcomp}
\usepackage{xcolor}
\usepackage{comment}
\usepackage{multirow}
\usepackage{tikz}

\newcommand\copyrighttext{%
  \footnotesize \textcopyright 2026 IEEE. Personal use of this material is permitted. Permission from IEEE must be obtained for all other uses, in any current or future media, including reprinting/republishing this material for advertising or promotional purposes, creating new collective works, for resale or redistribution to servers or lists, or reuse of any copyrighted component of this work in other works.}
\newcommand\copyrightnotice{%
\begin{tikzpicture}[remember picture,overlay]
\node[anchor=south,yshift=10pt] at (current page.south) {\fbox{\parbox{\dimexpr\textwidth-\fboxsep-\fboxrule\relax}{\copyrighttext}}};
\end{tikzpicture}%
}

\def\BibTeX{{\rm B\kern-.05em{\sc i\kern-.025em b}\kern-.08em
    T\kern-.1667em\lower.7ex\hbox{E}\kern-.125emX}}
\begin{document}

\title{On-Demand Instructional Material Providing Agent Based on MLLM for Tutoring Support
\thanks{This work was supported in part by JSPS KAKENHI Grant Numbers JP24K03052 and JP22K18006.}
}

\author{
\IEEEauthorblockN{Takumi Kato} 
\IEEEauthorblockA{
\textit{Nagoya Institute of Technology} \\
Nagoya, Aichi, Japan \\
tkato@ozlab.org}
\and
\IEEEauthorblockN{Masato Kikuchi} 
\IEEEauthorblockA{\textit{Nagoya Institute of Technology}\\
Nagoya, Aichi, Japan \\
kikuchi@nitech.ac.jp}
\and
\IEEEauthorblockN{Tadachika Ozono} 
\IEEEauthorblockA{\textit{Nagoya Institute of Technology}\\
Nagoya, Aichi, Japan \\
ozono@nitech.ac.jp}

}

\maketitle
\copyrightnotice{}

\begin{abstract}
Effective instruction in tutoring requires promptly providing instructional materials that match the needs of each student (e.g., in response to questions). In this study, we introduce an agent that automatically delivers supplementary materials on demand during one-on-one tutoring sessions. Our agent uses a multimodal large language model to analyze spoken dialogue between the instructor and the student, automatically generate search queries, and retrieve relevant Web images. Evaluation experiments demonstrate that our agent reduces the average image retrieval time by 44.4 s compared to cases without support and successfully provides images of acceptable quality in 85.7\% of trials. These results indicate that our agent effectively supports instructors during tutoring sessions.
\end{abstract}

\begin{IEEEkeywords}
Multimodal Large Language Model, Tutoring, Agent, Educational Technology
\end{IEEEkeywords}

\section{Introduction}
Tutoring, where instructors guide only a small number of students, is an effective learning method for students, as it allows instruction to be tailored to each student's proficiency and needs. Effective tutoring requires prompt provision of instructional materials that address each student's questions. However, owing to the diversity of student questions and the difficulty of preparing suitable materials in advance, instructors often have to quickly prepare the materials after receiving the questions. Although instructors may hand-draw diagrams, doing so is not suitable for drawing complex figures. In such cases, obtaining appropriate images from Web-based resources becomes a promising alternative. However, the method of manually entering search queries into a search engine or manually entering prompts into a generative AI for image retrieval has the drawback of interrupting the instructor's explanation. Based on the above, we need a system that automatically provides appropriate instructional materials while the instructor is answering questions.

We developed an on-demand(i.e., just-in-time, context-aware) material-providing agent that retrieves and displays images appropriate for answering student questions based solely on the dialogue audio exchanged between the instructor and the student. Our approach focuses on using Web images by employing a multimodal large language model (MLLM) to generate search queries from dialogue audio. Allowing a large language model (LLM) to generate images can lead to hallucinations. Moreover, we found that generating a single image using OpenAI's GPT-4o takes, on average, approximately 90 s. In contrast, our approach involves searching for images available on the Web, thereby avoiding hallucinations. Furthermore, by tracing the source of the retrieved images, it becomes possible to analyze the rationale behind the selection of a particular image. Additionally, because image search requires less time than image generation, our method is better suited to meeting the demand for on-demand image display. In our preliminary experiment, we found that instructor-role participants take over one minute on average to manually search for Web images they consider appropriate. This system reduces the effort and time an instructor would otherwise spend searching for images after a student's question. Consequently, instructors and students can focus on the instructional process, potentially improving overall educational outcomes. Our contributions to realizing our agent are as follows: (1) obtaining contextual information from dialogue audio to search for images suitable for question and answer; (2) generating search queries from contextual information; (3) selecting and providing instructional images retrieved from the Web; (4) ensuring that the agent operates in real time.

\section{Related Work}
Intelligent tutoring systems (ITS) are designed to facilitate personalized learning and have been the subject of extensive research \cite{ITS}\cite{AIITS}. A variant of ITS is the conversational tutoring system (CTS) \cite{Ruffle&Riley}. CTSs support learning by engaging in natural language dialogue with learners, and their benefits—particularly in inference tasks—have been demonstrated \cite{SystematicLiterature}. For example, AutoTutor \cite{AutoTutor} creates a learning environment in which students are guided through natural language conversation; through text-based interactions, it corrects student errors and answers questions. Our research contributes to these areas by focusing on reducing the cognitive load on instructors during one-on-one tutoring sessions.

Related research on agents that provide materials includes systems such as Auto-Presentation \cite{Auto-Presentation}, an automated presentation construction system that analyzes, summarizes, and correlates information from Internet-based knowledge sources based on a user-supplied topic. The main features of Auto-Presentation include understanding user requests, an interactive character agent, dynamic construction of presentation outlines, and functions such as Web search, information extraction, summarization, correlation, and agent script generation. However, that system extracts a topic from user input and then performs a predetermined search based on that topic. In contrast, our system aims to realize an educational support system that takes into account the contextual nuances of the interaction between the instructor and the student.

Kao et al. \cite{VisualizingDialogues} proposed a multimodal dialogue system for seamless photo sharing. Their approach generates relevant visual descriptors from dialogue information using an LLM, which are then used to search for photos within a specific image database. Although this study is not specifically targeted at the educational domain, it represents a novel technical approach to addressing the challenge of image provision based on dialogue context, thereby expanding the potential for autonomous agents to respond multimodally. However, this study requires a preconstructed image repository from a set of object lists and photographs in advance, which makes it difficult to apply to tutoring, where students have a wide range of questions and it is difficult to prepare materials in advance.

In the context of one-on-one tutoring sessions, where it is essential to provide appropriate supplementary materials in response to learners' questions, it is difficult to pre-prepare suitable materials for every possible inquiry. Therefore, our study adopts an approach that uses Web images as a database and conducts image searches accordingly. Retrieving supplementary materials from the Web requires generating search queries, a task for which ChatGPT has shown effectiveness \cite{ChatGPT}. The task of generating search queries from dialogue audio is a form of information extraction. In this field, techniques for recognizing named entities (e.g., people's names, organization names, locations, and numerical data) from text are essential. Hu et al. \cite{LLMNER} demonstrated that using LLMs can be effective for named entity extraction. Based on these results, our research employs an LLM to generate search queries.
\section{Challenges and Proposed Approach}\label{sec:problem_solution}
\subsection{Challenges}
Tutoring is characterized by the instructor's ability to focus on each student. There are situations where the instructor is explaining a concept and wishes to provide supplementary materials to support the explanation. In such cases, two main challenges arise. First, it is extremely difficult to prepare supplementary materials in advance for every possible scenario; the variety of student questions makes it nearly impossible to anticipate all needs. Even if the instructor is familiar with the curriculum, questions can arise about practice problems or topics beyond the standard school content. Second, when unexpected questions come up, the instructor faces a high cognitive load because multiple factors must be considered (e.g., identifying the student's main point of confusion, assessing the student's current understanding, and deciding on an appropriate explanation sequence). Additionally, preparing new supplementary materials on the fly may interrupt the explanation, potentially causing the student to lose concentration. Therefore, if appropriate supplementary materials can be obtained automatically, the instructor's burden can be alleviated.

When it comes to providing a diagram that has not been pre-prepared, possible approaches include: (1) hand-drawing the diagram, (2) manually searching for printed materials, (3) manually searching for appropriate images on the Web, and (4) generating images using generative AI. It is also possible to add hand-drawn annotations to the retrieved images. Hand-drawn drawings allow for interactive explanations of concepts but are not suitable for complex figures such as three-dimensional shapes or crystal structures. In contrast, approaches (2) and (3) tend to interrupt the explanation, as the instructor must shift focus to perform a manual search. In approach (4), the use of generative AI can also disrupt the instructor's explanation and may cause hallucinations in the generated images. Furthermore, image generation with generative AI is time-consuming and unsuitable for answering real-time questions. For instance, generating images with GPT-4o took approximately 90 s on average. To solve this problem, we aim to develop a system that provides necessary materials to the instructor without interrupting the dialogue. Specifically, our study focuses on automatically generating search queries for image retrieval from the dialogue audio exchanged between the instructor and the student using an LLM.

\subsection{On-Demand Instructional Material Providing Agent}

We introduce our material-providing agent, which can autonomously and instantly provide images suitable for instructional responses based solely on the dialogue audio, including question–answer exchanges between the instructor and the student. Instead of waiting for the instructor to finish listening to the student's question and then issuing an image search command, the agent continuously monitors the dialogue and autonomously provides materials without requiring explicit instructions from the instructor.

However, simply using an MLLM is not sufficient to realize this agent. During the instruction-tuning phase of training, an MLLM learns to respond to explicit user commands. That is, it is not trained to execute tasks that are not explicitly instructed, even if they can be inferred from multiparty dialogue. Therefore, if a student says, ``I don't understand …,'' the MLLM might try to answer the question, or if the instructor says, ``Calculate …,'' the MLLM might attempt the calculation—actions that may not be desired. This lack of control over its behavior makes direct employment of an MLLM challenging. Although an alternative approach involves the instructor verbally prompting the MLLM to retrieve relevant images in response to student questions, such interruptions in the dialogue are undesirable.

In this system, rather than automatically generating the necessary materials using generative AI, our approach reuses existing images available on the Web. To this end, we propose using MLLM to generate search queries from the dialogue audio. This system reduces the time and effort required for an instructor to search for images after a student's question, thereby allowing both the instructor and the student to focus on the instruction and potentially enhancing educational outcomes. The implementation of this autonomous material-providing agent presents several challenges: (1) obtaining contextual information from dialogue audio to retrieve images relevant to the question–answer exchange, (2) generating search queries based on that context, (3) selecting and providing instructional images retrieved from the Web, and (4) ensuring on-demand performance. To address these challenges, our approach consists of the following:
\begin{itemize}
    \item \textbf{Approach (1):} Utilize an MLLM to generate search queries from dialogue audio, which are then used to search for relevant Web images to support explanations. This addresses both Challenge 1 (obtaining contextual information) and Challenge 2 (generating search queries).
    \item \textbf{Approach (2):} Automate the entire process–from capturing the audio to providing the retrieved images–to address Challenge 4 (on-demand performance). For Challenge 3 (selecting and providing images), we initially adopted a simple method by providing the top-ranked search results.
\end{itemize}

\section{Implementation}\label{sec:implementation}
\subsection{Architecture}
\begin{figure}[t]
    \centering
    \includegraphics[width=0.5\textwidth]{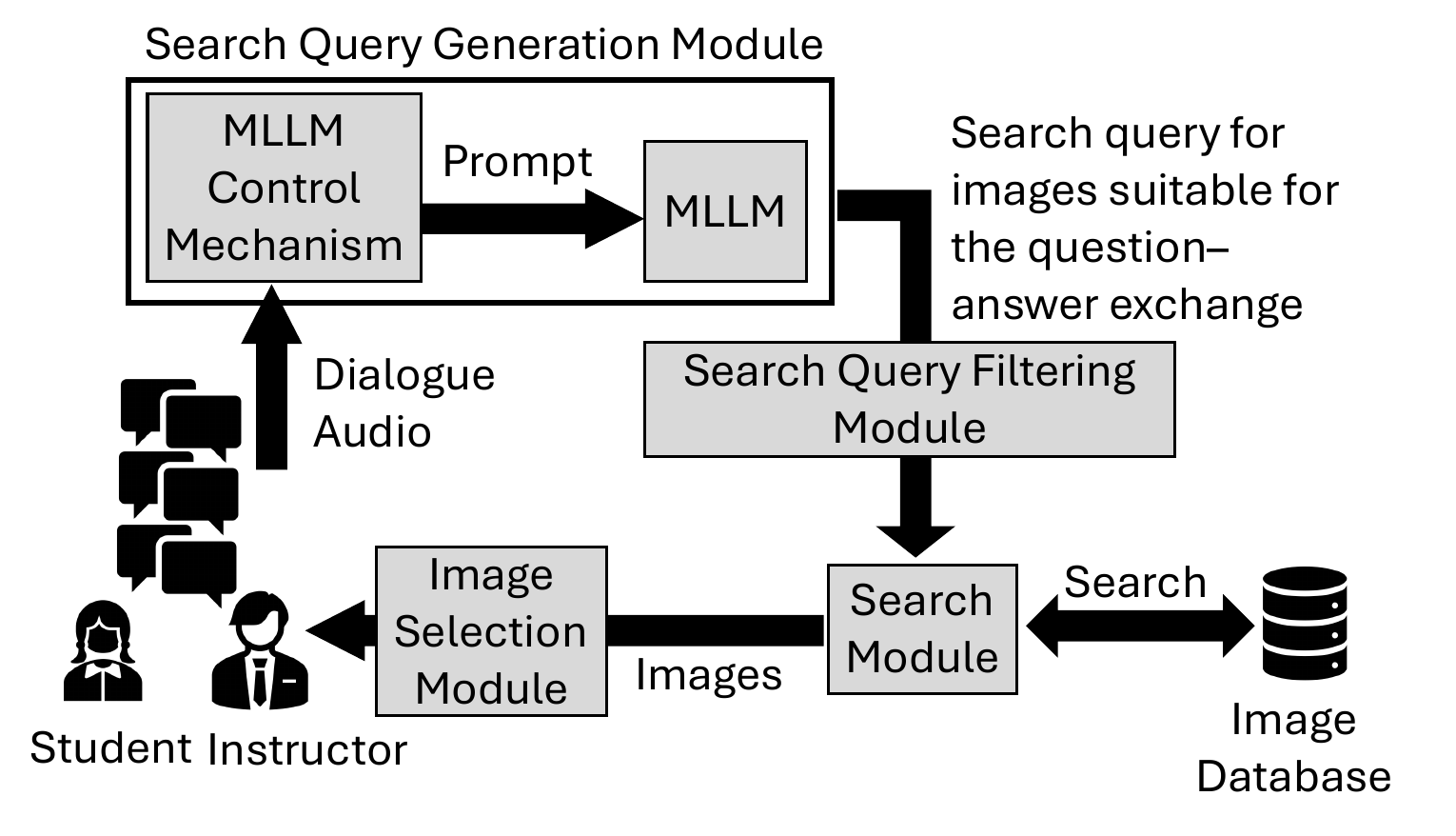}
    \caption{Proposed Method}
    \label{fig:system_ideal}
\end{figure}

As shown in Fig.~\ref{fig:system_ideal},
the proposed system is composed of four modules:
\begin{enumerate}
    \item \textbf{Search Query Generation Module:} Generates search queries from dialogue audio.
    \item \textbf{Search Query Filtering Module:} Filters out unnecessary or harmful search queries.
    \item \textbf{Search Module:} Performs image searches in the image database based on the search queries.
    \item \textbf{Image Selection Module:} Selects useful images from the search results.
\end{enumerate}
The processing flow is as follows. The search query generation module (1) is the core of our method; it generates search queries to retrieve images suitable for the question–answer exchange from the dialogue audio. The generated queries are then passed to the filtering module (2), which checks for unwanted queries (e.g., those that might retrieve harmful content). The search module (3) then uses the filtered queries to retrieve images from the image database. Finally, the image selection module (4) ranks and filters the retrieved images and provides them via an appropriate user interface.

In this study, we focused on the search query generation module (1) and developed a prototype system. In this prototype, the implementations for the search query filtering module (2) and the image selection module (4) in Fig.~\ref{fig:system_ideal} were kept minimal—specifically, no filtering was performed in (2), and (4) simply used the ranking provided by (3).

\subsection{Search Query Generation Module}

This module generates search queries using an MLLM based on the dialogue audio of the question–answer session between the instructor and the student. Specifically, the MLLM control mechanism within the search query generation module constructs a prompt and then instructs the MLLM to generate search queries.

For the MLLM control mechanism to extract the relevant context from the dialogue audio for the question–answer exchange, it is preferable to supply only the necessary portions of the audio to the MLLM. For instance, factors such as topic changes or the inclusion of unrelated audio segments may hinder understanding; therefore, selecting only the relevant portions is important. However, because this task is non-trivial, our prototype development has the following two limitations:
\begin{enumerate}
    \item The dialogue audio does not contain topic changes.
    \item The dialogue audio includes only the conversation between the instructor and the student currently of interest.
\end{enumerate}
Given these limitations, this study focuses primarily on evaluating the time required for image search and the quality of the images retrieved from dialogue audio. The design of the MLLM control mechanism, based on these assumptions, is described below.

\begin{figure}[t]
    \begin{center}
        \includegraphics[width=0.5\textwidth]{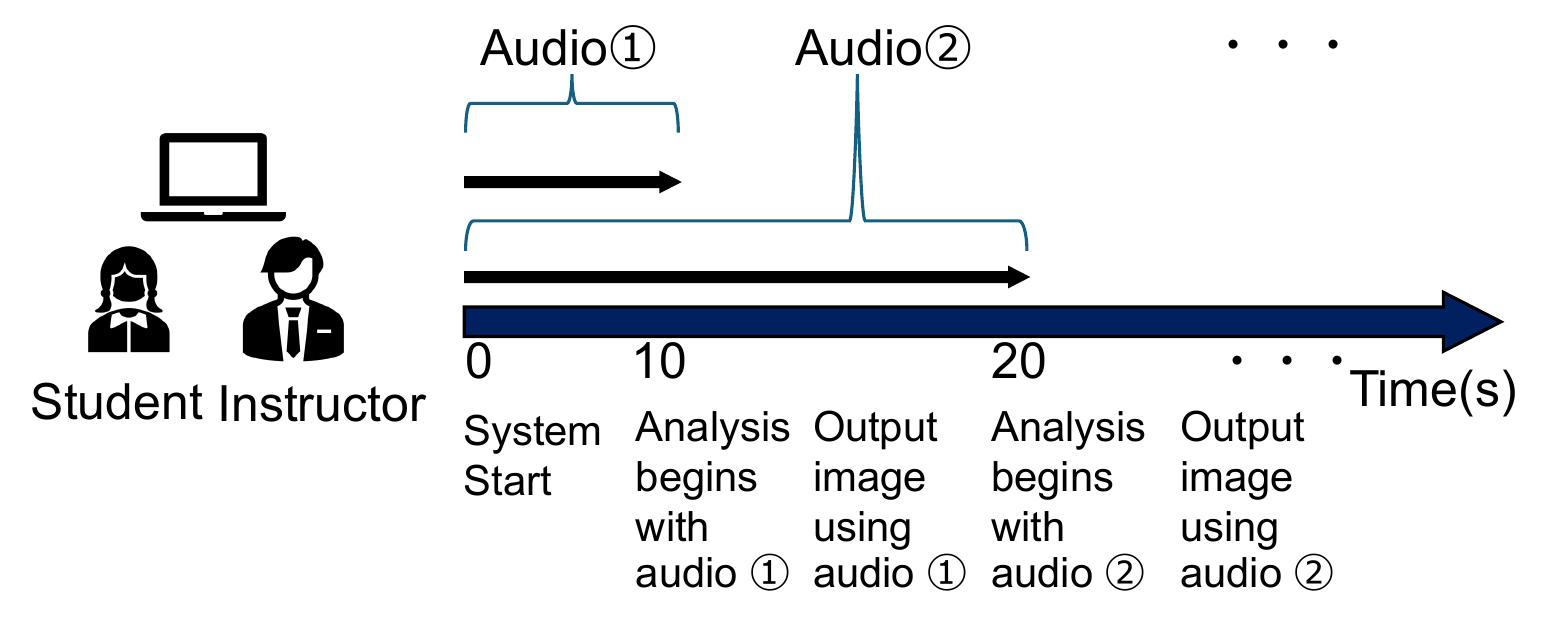}
    \end{center}
    \caption{Overview of the System Operation Process}
    \label{fig:system_process}
\end{figure}

The control mechanism operates at specific time intervals, as shown in Fig.~\ref{fig:system_process} , and utilizes the prompt shown in Fig.~\ref{fig:prompt} during its operation. Fig.~\ref{fig:system_process} illustrates the time intervals of the dialogue audio that are used for search query generation. For example, the timestamps 0, 10, and 20 represent the elapsed time (in seconds) from the start of the system (with the start defined as the moment when the student begins asking a question). The control mechanism instructs the MLLM to generate search queries every 10 s. At each interval, the MLLM is provided with a prompt along with the dialogue audio from the start time up to the current time (e.g., from 0–10 s at the 10-second mark, or 0–20 s at the 20-second mark). Given the assumption of no topic changes, incorporating a longer segment of dialogue audio is considered beneficial.

The prompt in Fig.~\ref{fig:prompt} begins by explaining that the accompanying audio file contains the dialogue related to the question–answer exchange. The script then provides instructions for generating search queries and finally specifies the required output format.

In our method, we use Google's ``Gemini 1.5 Flash'' as the MLLM. Considering the need for an on-demand search, processing speed was a critical factor in selecting the MLLM. Gemini 1.5 Flash is a Transformer decoder model optimized for computational efficiency, enabling rapid inference. Moreover, it achieves low latency by transferring knowledge from Gemini 1.5 Pro through online distillation techniques. According to Reid et al. \cite{Gemini}, the Gemini 1.5 model achieves near-perfect recall on long-context search tasks involving text, video, and audio—surpassing techniques in long-form question answering (QA), extended video QA, and long-form automatic speech recognition (ASR). In tests evaluating how reliably the models can recall information in distracted contexts, Gemini 1.5 Pro and Gemini 1.5 Flash achieved nearly perfect recall (\>99\%) up to multiple millions of tokens.


\section{Evaluation Experiments}\label{sec:evaluation}

In this section, we describe the evaluation experiments for the autonomous material-providing agent. We evaluated our proposed method in two aspects:
\begin{enumerate}
    \item \textbf{Output Time Evaluation (Experiment 1):} Measuring the time required for the system to autonomously output images useful for explanations based on the dialogue audio between the instructor and the student.
    \item \textbf{Output Image Evaluation (Experiment 2):} Assessing whether the system's output is comparable to that obtained through manual image search.
\end{enumerate}
Specifically, in Experiment 1, we measured the time required for the system to obtain an image of comparable quality to that retrieved via manual Web search. In Experiment 2, participants used the system and we observed whether they were satisfied with the system's output or opted to perform a manual Web search before the system provided a satisfactory result. In both experiments, the tests were conducted in pairs with laboratory students as participants–one acting as an instructor and the other as a student. The instructor prepared in advance on a given topic and then responded to the student's questions. The selected topics and questions were designed to require supplementary materials for explanation. In this experiment, the instructors were provided books related to the topic—in this case, earth science, allowing them to assess the relevance of the retrieved images. The following subsections detail each experiment. What we ideally want to measure is the time from when the instructor wants an image until it is displayed; however, this is difficult to measure objectively. Therefore, in these experiments, the time was measured by setting the start of the student's question as 0 s.

\begin{figure}[tb]
    \centering
    \fbox{
        \begin{minipage}{0.45\textwidth}
            {\normalsize
            This audio file is a recording of a student asking a question to the instructor.\\
            Please output search queries to retrieve Web images that focus on the specific issues of the student and facilitate understanding of the content.\\
            However, output exactly five search queries, ordered from most to least effective.\\[1ex]
            For example, the following search queries were generated for a student who did not understand a problem involving ``determining the increase and decrease, finding the extrema and drawing the graph of a cubic function.''\\
            Please follow this format exactly and output only the search queries.\\[1ex]
            Cubic Function Increase-Decrease Graph, Cubic Function Extrema Graph Drawing Method, Derivative Extrema, Increase-Decrease Table Graph, Derivative Increase-Decrease Table
            }
        \end{minipage}
    }
    \caption{Prompt Used to Retrieve Image Search Query}
    \label{fig:prompt}
\end{figure}

\subsection{Experiment 1: Output Time Evaluation}\label{sec:evaluation1}

The objective of Experiment 1 was to measure the time required for the system to obtain an image of quality comparable to that obtained by manual Web search. To establish a baseline, the instructor first performed a manual Web search to retrieve an image for the explanation (without system support). In the manual search case, the instructor listened to the student's question, formulated search queries, and engaged in a trial and error process until an appropriate image was obtained, known as the ``Web search image''. The time elapsed from when the student initiated the question to when an appropriate image was obtained was recorded as the Image Output Time (without support). As illustrated in Fig.~\ref{fig:system_process}, the moment the student asked the question was set to 0 s (as in Experiment 2). The same question was then asked again, this time with the system active. In this case, Image Output Time (with support) was measured from the start of the student's question to the point at which the participant judged that the system had output an image of equal or higher quality than the ``Web search image''. If the system failed to output an appropriate image, the result was marked as a failure.

Three participants took part in this experiment, all of whom played the role of instructor once. Most participants also took on the role of the student when others acted as instructors. This experiment was carried out using seven topics, each tested by a pair of instructors and students.

It is worth noting that the system's output image did not need to exactly match the Web search image. If the participant deemed the system-generated image to be of comparable or superior quality, the instructor could proceed with the explanation. Although this may introduce bias, addressing this issue is left for future work. Because the Web search image serves as a baseline for evaluation, an exact match is not essential. The reason for this setting is that Web images may contain multiple figures with similar contents, and the primary goal is to demonstrate the effectiveness of the system by outputting an image with equivalent or superior explanatory power, rather than one that perfectly matches the Web search image.

\subsection{Results of Experiment 1}\label{sec:evaluation1_result}
\begin{table}[tb]
    \caption{Results of Experiment 1: Comparison of With and Without System Support}
    \renewcommand{\arraystretch}{1.2}
    \label{table:evaluation1}
    \centering
    \begin{tabular}{|c|l|r|r|}
    \hline
    \multicolumn{1}{|c|}{\multirow{2}{*}{Subject}} & \multicolumn{1}{c|}{\multirow{2}{*}{Topic}} & \multicolumn{2}{c|}{Image Output Time [sec]} \\\cline{3-4}
& & \multicolumn{1}{c|}{Without Support} & \multicolumn{1}{c|}{With Support} \\\hline
    \hline
    \multicolumn{1}{|c|}{\multirow{4}{*}{A}} & Volcanic Eruption 1 & 50 & 16\\
& Volcanic Eruption 2 & 108 & 34\\
& Igneous Rock & 94 & 22\\
& Magma Crystallization & 60 & 22\\\hline
    \multicolumn{1}{|c|}{\multirow{2}{*}{B}} & Sea Fog & 73 & N/A \\
& El Ni\~{n}o & 52 & 14\\\hline
    \multicolumn{1}{|c|}{\multirow{2}{*}{C}} & F\"{o}hn Phenomenon & 30 & 20\\
& Spring Snow & 108 & N/A\\\hline
    \hline
    \multicolumn{2}{|c|}{AVG$\pm$SD (excluding N/A)} & 65.7$\pm$29.4 & 21.3$\pm$7.0\\\hline
    \end{tabular}
\end{table}
The results of Experiment 1 are shown in TABLE~\ref{table:evaluation1}. This table lists the Image Output Times for each topic, both without support and with support. In the supported case, the average time was reduced by 44.4 s compared to the unsupported case. (For the topic ``Volcanic Eruption,'' two images were used; hence, the time until the first image is labeled ``Volcanic Eruption 1'' and the time until the second image is labeled ``Volcanic Eruption 2.'') An ``N/A'' in the Image Output Time (with support) indicates that an image comparable to the Web search image was not outputted. Our interpretation of these cases is discussed in Section~\ref{sec:discussion}. When computing averages and standard deviations, topics labeled as ``N/A'' were excluded. In this experiment, none of the system's outputs was judged to be of higher quality than the Web search images. However, for the topics in which the system did produce an acceptable image, the image was identical to or comparable to the Web search image, and the time to obtain the result was shortened by an average of 44.4 s.

\subsection{Experiment 2: Output Image Evaluation}\label{sec:evaluation2}

In Experiment 2, we evaluated whether the system's output was comparable to that obtained through manual image search. As in Experiment 1, the instructor conducted a question-and-answer session while using the system. If the instructor was dissatisfied with the system's output images, they were allowed to conduct a manual Web search and use the retrieved image for their explanation. Under these conditions, a trial was considered successful if the instructor used the system's output image in their explanation. In contrast, if the instructor opted to use an image retrieved by manual Web search instead, the trial was marked as a failure.

The experimental setup was again designed so that the student asked questions requiring supplementary visual materials for the instructor's response. To avoid forcing the instructor to strictly adhere to the system and to prevent bias, the participants were not informed in advance of the experiment's purpose; they were simply told that their usage would be observed. Seven participants took part in this experiment, all of whom played the role of instructor once. Most participants also played the role of the student while others played the role of the instructor. Because each instructor–student pair was assigned two topics, a total of 14 topics were tested.

\subsection{Results of Experiment 2}\label{sec:evaluation2_result}
Experiment 2 was successful in 12 out of the 14 topics, resulting in a system success rate of 85.7\%. These results suggest that the images provided by the system were generally satisfactory to instructors for supplementing their explanations.

Considering the results of Experiment 1 and Experiment 2 together, although the system's image search accuracy sometimes lagged behind manual search, it still managed to retrieve relevant images that supported the instructor's explanations. Furthermore, the system offered the added benefit of reducing the time required for image retrieval.

\section{Discussion}\label{sec:discussion}
\subsection{Regarding Experiment}
TABLE~\ref{table:evaluation1} shows that in many cases the system was able to output images that were comparable to the Web search images; however, this was not true for all topics. Moreover, no images output by the system were judged to be superior to the Web search images, indicating that the system's image retrieval capabilities still do not yet fully match those of human search.

Furthermore, when comparing the Image Output Times between the unsupported and supported cases, the data in TABLE~\ref{table:evaluation1} indicate that for topics where the system produced an acceptable image, the supported case not only had a lower average time but also a smaller standard deviation. This suggests that the system not only reduces the average time required for image retrieval but also stabilizes retrieval times. In addition, because the images are output automatically, the time the instructor would otherwise spend on manual searches can be devoted to the explanation; thus, the practical benefit may exceed the measurable time savings.

In cases where the system did not output an image comparable to the Web search image, the images were still related to the topic. Although they were not optimal as supplementary materials, they could still be useful for the explanation, meaning that the instructor might still be able to use the system's output. This is because the requirement for the system is not to produce the optimal image 100\% of the time, but rather to output an image that the instructor deems acceptable for explanation.

In a preliminary experiment preceding Experiment 2, a participant conducting manual Web searches still failed to find a suitable image. For instance, participants' search queries included terms such as ``volcanic clastic material,'' ``volcanic gravel,'' and ``volcanic bomb volcanic gravel volcanic ash difference.'' In contrast, when using the corresponding audio, our system generated search queries such as ``volcanic clastic material difference,'' ``volcanic clastic material classification image,'' and ``volcanic clastic material magma.'' The participant's queries were simple word lists, whereas the system included terms like ``difference'' and ``classification.'' Although in some cases human-generated queries were more sophisticated, this finding suggests that our system has the potential to assist in formulating effective search queries.

\subsection{Future Directions}
In our approach, we used publicly available images on the Web as the image database; however, this is not a necessity. By employing an image database specialized for a particular field, it may be possible to achieve more accurate image searches. For example, if a tutoring center has its own textbooks or dedicated materials, an image database could be constructed based on those data. Furthermore, there is room for improvement in how the images are presented. Although our current method outputs diagrams relevant to the topic, alternative approaches might involve generating images designed for instructor annotation.

Although our system is intended for educational support, it could be adapted to other fields with appropriate prompt modifications. For instance, during a presentation, the system could display images relevant to the topic being discussed, thereby enhancing communication. In this way, our system may prove useful in various contexts where conversational interaction plays a central role.

In this study, we adopted an LLM-based implementation designed to be applicable in a range of academic fields. Further development could lead to systems that automatically generate presentation slides in sync with spoken content. However, this general-purpose design may limit output quality when compared to systems fine-tuned for specific domains. Consequently, future work may involve developing both a general-purpose system and specialized versions for particular fields.

Finally, our current implementation assumes that the dialogue audio does not include topic changes and contains only the conversation relevant to the explanation. To further develop this system, it will be necessary to handle situations where these assumptions do not hold. In particular, addressing topic transitions may involve developing algorithms that discard older audio at appropriate intervals or detect topic-change boundaries, all without compromising on-demand performance. Furthermore, reducing the operational cost of using the LLM during long-term deployment remains a challenge for future research.

\section{Conclusion}
In this study, we developed an on-demand material-providing agent designed to support instructors during tutoring sessions. This agent autonomously retrieves and presents relevant images based solely on the audio dialogue between the instructor and the student, without requiring manual search requests. A key feature of the system is its ability to periodically generate context-specific search queries using an MLLM and deliver images retrieved based on those queries, allowing the user to obtain the necessary images without interrupting the dialogue.

Evaluation experiments demonstrated that the system reduced the time required to obtain a relevant image by an average of 44.4 s compared to the unsupported case. Furthermore, in 85.7\% of trials, the system's output was deemed acceptable for instructional use. These results indicate that the system is useful in reducing image search time.

Future work includes addressing challenges such as topic changes, extending the system's operation over long-term operation, and devising methods to reduce experimental bias. In addition to alleviating the burden on tutors, this system can also be applicable in ad hoc situations, such as presentations or product demonstrations, where additional materials are required.


\bibliographystyle{unsrt}
\bibliography{ref}
\end{document}